\documentclass[final,10pt,twocolumn,a4paper,conference]{IEEEtran}

\usepackage{amsmath,amsfonts,amssymb,amsbsy,url,verbatim}
\usepackage{mathtools}
\usepackage{times}
\usepackage[english]{babel}
\usepackage[dvips]{graphicx}
\usepackage{bm}
\usepackage{comment}
\usepackage{stfloats}
\usepackage{cite}
\usepackage{psfrag}
\usepackage{dsfont} 
\usepackage{url}
\usepackage{wrapfig}
\usepackage{algorithm}
\usepackage{algcompatible}
\usepackage{boxedminipage}
\usepackage{multicol}

\newcommand{\bi}{\begin{itemize}}
\newcommand{\ei}{\end{itemize}}
\newcommand{\ben}{\begin{enumerate}}
\newcommand{\een}{\end{enumerate}}
\newcommand{\bc}{\begin{cases}}
\newcommand{\ec}{\end{cases}}
\newcommand{\bd}{\begin{description}}
\newcommand{\ed}{\end{description}}

\newcommand{\be}{\begin{equation}}
\newcommand{\ee}{\end{equation}}
\newcommand{\bea}{\begin{eqnarray}}
\newcommand{\eea}{\end{eqnarray}}

\newcommand{\back}{\!\!\!\!\!}

\IEEEoverridecommandlockouts

\begin{document}

\title{Towards an Appropriate Receiver Beamforming Scheme for Millimeter Wave Communication: \\
A Power Consumption Based Comparison\vspace{-4mm}}

\author{Waqas Bin Abbas, Michele Zorzi \thanks{The work of Michele Zorzi was partially supported by NYU Wireless.}\\
    \IEEEauthorblockA{Department of Information Engineering, University of Padova, Italy \vspace{-4.5mm}}\\
    \IEEEauthorblockA{E-mail: \texttt{\small \{waqas,zorzi\}@dei.unipd.it} 
}}

\maketitle

 \begin{abstract}
At millimeter wave (mmW) frequencies, beamforming and large antenna arrays are an essential requirement to combat the high path loss for mmW communication. 
Moreover, at these frequencies, very large bandwidths are available to fulfill the data rate requirements of future wireless networks.
However, utilization of these large bandwidths and of large antenna arrays can result in a high power consumption which is an even bigger concern for mmW receiver design.
In a mmW receiver, the analog-to-digital converter (ADC) is generally considered 
as the most power consuming block.   
In this paper, primarily focusing on the ADC power, we analyze and compare the total power consumption of the complete analog chain for Analog, Digital and Hybrid beamforming (ABF, DBF and HBF) based receiver design.
We show how power consumption of these beamforming schemes varies with a change in the number of antennas, the number of ADC bits ($b$) and the bandwidth ($B$). 
Moreover, we compare low power (as in \cite{OrhanER15_PowerCons}) and high power (as in  \cite{Alkhateeb15_SwvsPS}) ADC models, and show that for a certain range of number of antennas, $b$ and $B$, DBF may actually have a comparable and lower power consumption than ABF and HBF, respectively.
In addition, we also show how the choice of an appropriate beamforming scheme depends on the signal-to-noise ratio regime.  

 \end{abstract}
 
\section{Introduction}
\label{sec:Rlt_wk}
The millimeter wave (mmW) spectrum (30-300 GHz), where a very large bandwidth is available, is considered as a prime candidate to fulfill the data rate requirements of future broadband communication \cite{KhanF_mmWave}. 
However, communication at these frequency bands exhibits high pathloss.
To overcome this high pathloss, spatial beamforming using large antenna arrays is considered as an essential part of a mmW communication system.

Analog, Hybrid and Digital beamforming (ABF, HBF and DBF) are the beamforming schemes being considered \cite{SM_BF_2014}.
Traditionally, digital beamforming is a popular choice, as it provides the advantages of digital processing techniques (such as multi-user communication, interference cancellation, formation of multiple simultaneous beams, etc), thanks to the use of a separate RF chain and analog-to-digital converter (ADC) per antenna element \cite{SM_BF_2014}. 
However, the utilization of many antennas and the large bandwidths at mmW result in a high power consumption, which generally makes DBF less desirable for power constrained mmW receiver design \cite{mmW_BF_2014}.
To reduce the power consumption, a hybrid scheme which performs beamforming in both the analog and the digital domain with a reduced number of RF  chains (at the cost of lower flexibility than DBF) is often presented as an attractive choice \cite{mmW_BF_2014}.
Moreover, ABF has the least power consumption and is an attractive beamforming choice whenever the advantages of digital processing techniques are not required.    

In this paper, we argue that the general perception regarding the high power consumption of DBF is not always true when the total power consumption of the receiver is considered.
Rather, there is a certain range of system bandwidth (suitable to fulfill the requirements of certain key functionalities of future wireless communication) and ADC resolution (to avoid any significant performance loss) for which DBF results in a power consumption lower than HBF and comparable to ABF while providing the flexibility of digital processing, 
which makes DBF an attractive candidate for mmW receiver design.

\subsection{Related Work}
\label{sec:rlt_work}  
Recently, energy efficient designs have been studied, particularly focusing on how the system capacity varies as a function of the ADC resolution.
In \cite{Madhow_1bitADC}, an exact nonlinear quantizer model is utilized to evaluate the optimal capacity for a 1-bit ADC.
In \cite{Opt_ADC_Res}, considering a MIMO channel and an additive quantization noise model (AQNM, an approximate model for ADCs), a joint optimization of ADC resolution and number of antennas is studied.
In a recent work \cite{OrhanER15_PowerCons}, the authors studied how the number of ADC bits $b$ and the bandwidth (sampling rate) $B$ of ADCs affect the total power consumption for ABF and DBF based receivers.
They studied the optimal $b$ and $B$ which maximize the capacity for ABF and DBF only for low power receiver design while also showing that DBF with similar power budget to ABF may achieve a higher rate than ABF when the channel state information is available at the transmitter. 
Recently, in \cite{Alkhateeb15_SwvsPS}, to further reduce the power consumption of HBF a switch based architecture is proposed, where at a particular instant only a reduced set of antennas (equal to the number of RF chains) is selected and connected to the RF chains. 
However, the reduction in the number of antennas also reduces the antenna array gain.

To the best of our knowledge, there is no previous work that compares the \emph{total} power consumption of ABF, HBF and DBF based receiver design for different values of the number of receive antennas $N_{ANT}$, the number of ADC bits $b$, and the bandwidth $B$.
In this paper, we provide a comprehensive comparison of total power consumption of different beamforming schemes while considering both high power \cite{Alkhateeb15_SwvsPS} and low power \cite{OrhanER15_PowerCons} ADC models.
Moreover, we discuss the relationship of the signal-to-noise ratio (SNR) with the ADC resolution and how it affects the choice of an appropriate beamforming scheme. 

\subsection{Our Contribution}
\label{ssec:Our_app}

In mmW receiver design, the ADC is usually considered to be the most power hungry block.
In general, DBF, which requires a number of ADCs ($N_{ADC}$) equal to twice the number of receive antennas (separate ADCs for each inphase and quadrature phase signal) is typically assumed to result in maximum power consumption, while ABF with $N_{ADC} = 2$ is the least power consuming scheme.
On the other hand, HBF requires $N_{ADC} = 2\times N_{RF}$ where $N_{RF} \leq N_{ANT}$ is the number of RF chains, and generally is assumed to have a lower power consumption than DBF. 
  \begin{figure}
        \centering
        \psfrag{Access Error Probability }{\ \ \ \ \ \ \ \ $P_{AccEr}$}
        \psfrag{Tx-Rx Distance }{\back\ $Tx-Rx Distance$}
        \scalebox{0.8}[0.6]{\includegraphics[width=\columnwidth]{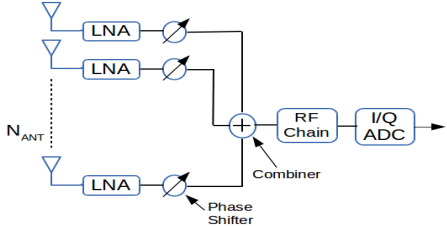}}\vspace{-4mm}
        \caption{\protect\renewcommand{\baselinestretch}{1.25}\footnotesize Analog Beamformer . }
\label{fig:ABF}
\end{figure} 
  \begin{figure}
        \centering
        \psfrag{Access Error Probability }{\ \ \ \ \ \ \ \ $P_{AccEr}$}
        \psfrag{Tx-Rx Distance }{\back\ $Tx-Rx Distance$}
        \scalebox{0.8}[0.6]{\includegraphics[width=\columnwidth]{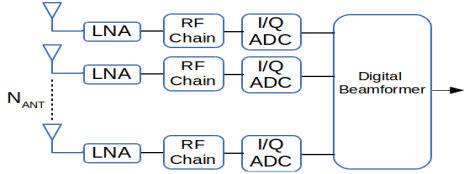}}\vspace{-4mm}
        \caption{\protect\renewcommand{\baselinestretch}{1.25}\footnotesize Digital Beamformer . }
\label{fig:DBF}
\end{figure}

The commonly accepted conclusion that DBF suffers from high power consumption is the result of implicitly assuming the use of high resolution and wide band ADCs, that therefore dominate the overall power budget.
However, the power consumption of an ADC is directly proportional to the number of quantization levels and to the sampling rate.
In addition, for different beamforming schemes, a power consumption comparison only based on ADCs can result in a different outcome with respect to what would be obtained when considering the total power consumption, especially when the resolution and/or the bandwidth of the ADC are not large.  
In this paper, we compare the total power consumption ($P_{Tot}$) of ABF, DBF and HBF by considering a low power ADC (LPADC) and a high power ADC (HPADC) models,
and for different values of $N_{ANT}$, $B$ and $b$. Our results show that

\begin{itemize}
\item $P_{Tot}$ for all beamforming schemes increases with an increase in $N_{ANT}$, $b$ or $B$;
\item for fixed $N_{ANT}$ and $B$, there is a maximum number of bits ($b^*$) up to which DBF is more energy efficient than HBF;  
\item for fixed $N_{ANT}$ and $b$, there is a maximum bandwidth $B^*$ up to which DBF  is more energy efficient than HBF; 
\item DBF always has higher power consumption than ABF, however, for small $b$ and $B$ the difference is relatively small and therefore, in those configurations, DBF may still be an attractive option, also in view of the much greater flexibility provided by digital processing;
\item in comparison to HBF, if the ratio of $N_{RF}$ and $N_{ANT}$ remains constant, $b^*$ and $B^*$ for DBF both increase with an increase in $N_{ANT}$;
\item the variation in $b$ has a more significant effect at high SNR than at low SNR.  
\end{itemize} 

Finally, the choice of the appropriate beamforming scheme depends not only on the total power consumption but also on the SNR regime. Although the qualitative trends among different beamforming schemes are rather predictable, the precise quantification of these relationships and the results presented in this paper are useful to precisely characterize the regimes where the various beamforming options are to be preferred.

\begin{figure}
        \centering
        \scalebox{0.85}[0.75]{\includegraphics[width=\columnwidth]{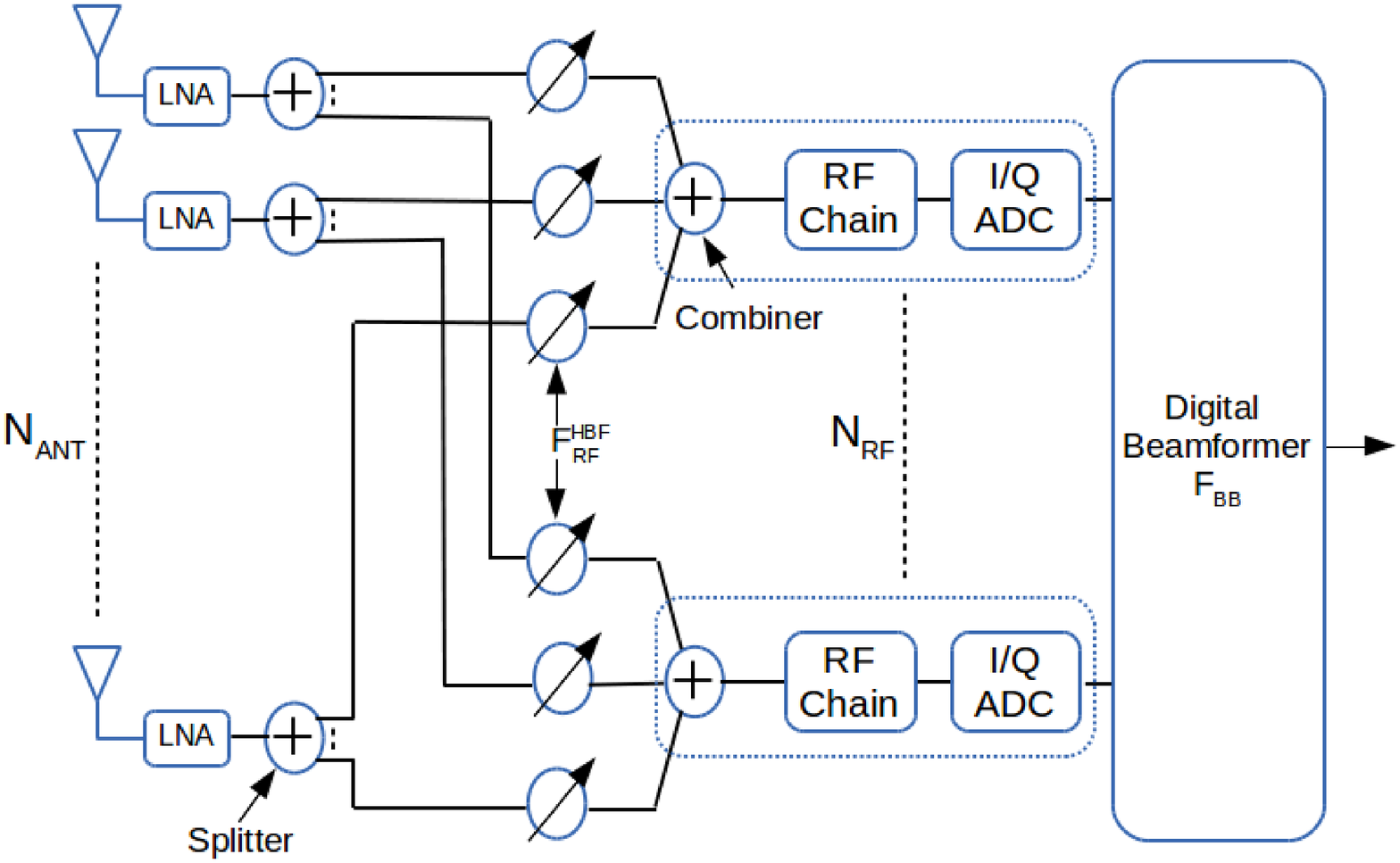}}\vspace{-4mm}
        \caption{\protect\renewcommand{\baselinestretch}{1.25}\footnotesize Hybrid Beamformer.}
\label{fig:HBf}
\end{figure}

\section{System Model} 
\label{Sec:Sys_mod}
Common mmW receiver architectures for ABF, DBF and HBF are shown in Figures \ref{fig:ABF}, \ref{fig:DBF} and \ref{fig:HBf}, respectively.
The total power consumption $P_{Tot}$ of these beamforming schemes can be evaluated as 
\begin{equation}
   P_{Tot}^{ABF} =  N_{ANT}(P_{LNA}+P_{PS}) + P_{RF} + P_C+ 2P_{ADC}
   \label{eq:ABF}
\end{equation}  
\begin{equation}
\begin{split}
   P_{Tot}^{HBF} = &\ N_{ANT}(P_{LNA} + P_{SP} + N_{RF}P_{PS})\\			& + N_{RF}(P_{RF}+P_C + 2P_{ADC} )
\end{split}
\label{eq:HBF}   
\end{equation}
\begin{equation}
   P_{Tot}^{DBF} =  N_{ANT}(P_{LNA} + P_{RF} + 2P_{ADC})
   \label{eq:DBF}
\end{equation} 
\normalsize
where $P_{RF}$ represents the power consumption of the RF chain and is given by
\begin{equation}
   P_{RF} =  P_M + P_{LO} + P_{LPF} + P_{BB_{amp}}
   \label{eq:Prf}
\end{equation}
\normalsize
and $P_{LNA}$, $P_{PS}$, $P_C$, $P_M$, $P_{LO}$, $P_{LPF}$, $P_{BB_{amp}}$, $P_{ADC}$, and $P_ {SP}$ represent the power consumption of low noise amplifier (LNA), phase shifter, combiner, mixer, local oscillator, low pass filter, baseband amplifier, ADC, and splitter, respectively. 
In our analysis, 
the power consumption of all other components except the ADC is considered independent of the system bandwidth, whereas
$P_{ADC}$ increases linearly with $B$ and exponentially with $b$ \cite{ADC_b_B}. Therefore, considering Nyquist sampling rate, $P_{ADC}$ in terms of $B$ and $b$ is given by
\begin{equation}
\begin{split}
   P_{ADC} =  cB2^b =  cBR
   \label{eq:P_ADC}
\end{split}
\end{equation} 
\normalsize 
where $c$ is the energy consumption per conversion step, and $R = 2^b$ is the number of quantization levels of the ADC.

\subsection{$P_{Tot}$ Comparison}
\label{ssec:Ptot_Comp}
\begin{figure}
        \centering
		\psfrag{Total Power Consumption (Watts)}{\ \ \ \ \ \ \ \ \ \ $P_{Tot}$ (Watts)}
        \psfrag{Number of ADC bits}{\ \ \ \ \ \ \ \ \ \ $b$}
        \psfrag{Bandwidth = 100 MHz}{\ \ \ $B = 100$ MHz}
        \scalebox{0.9}[0.65]{\includegraphics[width=\columnwidth]{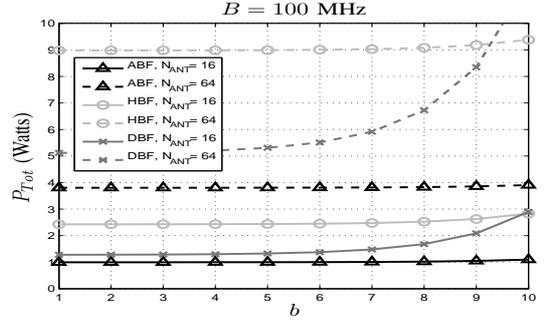}}\vspace{-4mm}
        \caption{\protect\renewcommand{\baselinestretch}{1.25}\footnotesize $P_{Tot}$ for different beamforming schemes vs $b$ for $B = 100$ MHz and $N_{ANT} = 16, 64$. }
\label{fig:AHDBF_100M}
\end{figure}

\begin{figure}
        \centering
		\psfrag{Total Power Consumption (Watts)}{\ \ \ \ \ \ \ \ \ \ $P_{Tot}$ (Watts)}
        \psfrag{Number of ADC bits}{\ \ \ \ \ \ \ \ \ \ $b$}
        \psfrag{Bandwidth = 1 GHz}{\ \ \ $B = 1$ GHz}
        \scalebox{0.9}[0.65]{\includegraphics[width=\columnwidth]{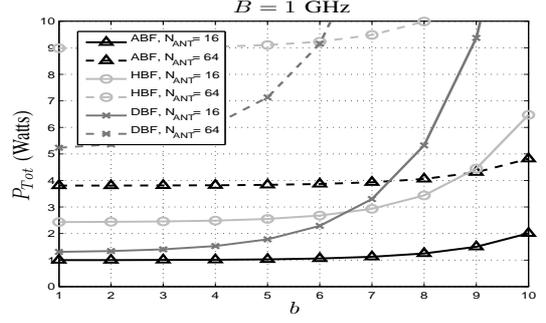}}\vspace{-4mm}
        \caption{\protect\renewcommand{\baselinestretch}{1.25}\footnotesize $P_{Tot}$ for different beamforming schemes vs $b$ for $B = 1$ GHz and $N_{ANT} = 16, 64$. }
\label{fig:AHDBF_1G}
\end{figure}

A comparison of $P_{Tot}^{ABF}$, $P_{Tot}^{HBF}$ and $P_{Tot}^{DBF}$ is shown in Figures \ref{fig:AHDBF_100M} 
and \ref{fig:AHDBF_1G} for $B$ equal to 100 MHz 
and 1 GHz, respectively.
In these plots, $N_{ANT}$ is set to 16 and 64, $b$ is varied from 1 to 10, and $N_{RF} = 4$ for HBF.
Moreover, $P_{LNA} = 39$ mW, $P_{PS} = 19.5$ mW, $P_M = 16.8$ mW, \cite{Rx_Pow_LNA_PS_C}, \cite{Rx_Pow_60GHz} $c = 494$ fJ  \cite{OrhanER15_PowerCons}, $P_{LO} = 5$ mW, $P_{LPF} = 14$ mW, $P_{BB_{amp}} = 5$ mW \cite{Alkhateeb15_SwvsPS} and $P_{SP} = 19.5$ mW.
Note that the results shown in Figures \ref{fig:AHDBF_100M} and \ref{fig:AHDBF_1G} are for the LPADC considered in \cite{OrhanER15_PowerCons}\footnote{Similar results can be obtained by considering HPADC (with $c \approx 12.5$ pJ) as in \cite{Alkhateeb15_SwvsPS}, which results in a reduced range of $b$ or $B$ for which DBF has a lower power consumption than ABF or HBF.  
We will mention the range of $b$ and $B$ for HPADC whenever necessary.}.

In Figures \ref{fig:AHDBF_100M} 
and \ref{fig:AHDBF_1G}, results show that $P_{Tot}$ increases with an increase in $N_{ANT}$, $B$ or $b$, as expected.
Firstly, note that ABF consumes the least power for every configuration. 
Secondly, DBF always has some configuration for which it has a lower power consumption than HBF. 
This is because $P_{ADC}$ increases exponentially with $b$, and therefore for small $b$ there is no significant power consumption due to $P_{ADC}$ with respect to the other components in Eq. (\ref{eq:DBF}). 
Moreover, at low $b$, the power consumption of additional components in HBF, e.g., phase shifters, becomes dominant and therefore HBF may even result in a higher power consumption than DBF. 
Note that the value of $b$ which results in a lower $P_{Tot}^{DBF}$ in comparison to $P_{Tot}^{HBF}$ (for fixed $N_{RF}$) decreases with an increase in $N_{ANT}$ and $B$.
For instance, for $N_{ANT} = 64$ and with $B = 1$ GHz and $B = 100$ MHz, $P_{Tot}^{DBF}$ is less than $P_{Tot}^{HBF}$ up to 6 bits and 9 bits, respectively.
Moreover, similar results obtained by considering an HPADC model \cite{Alkhateeb15_SwvsPS} (not shown here), show that $P_{Tot}^{DBF}$ always results in a higher power consumption than $P_{Tot}^{ABF}$ for the configurations used in Figure \ref{fig:AHDBF_100M}
and \ref{fig:AHDBF_1G}. 
However, DBF results in a lower power consumption than HBF for $B = 100$ MHz and $B = 1$ GHz and with $N_{ANT} = 16$ only for a range of $b$ up to 5 and 2, respectively. A further discussion on the impact of the number of bits is given in Section-\ref{sec:bits_SNR}.

We next provide analytical formulas to identify $B^*$ and $b^*$ for which $P_{Tot}^{DBF}$ is similar to $P_{Tot}^{HBF}$, for a general $N_{ANT}$. This is useful to properly characterize the regions in which DBF is to be preferred over the HBF alternative.

\subsection{Evaluation of $b^*$ and $B^*$}
\label{Sec:optimal_b_B}
We now compare DBF with HBF, and evaluate the maximum number of bits $b^*$ and the maximum bandwidth $B^*$ which satisfy the condition that 
$P_{Tot}^{DBF} \leq P_{Tot}^{HBF}$.

To find the values of $b^*$ and $B^*$ that result in the same total power consumption for HBF and DBF we first evaluate the intersection point of Eqs. (\ref{eq:HBF}) and (\ref{eq:DBF}).
This gives the following result
\small
\begin{equation}
\begin{split}
(N_{ANT} - N_{RF})&P_{RF} + 2(N_{ANT} - N_{RF})P_{ADC} = \\
 &N_{ANT}N_{RF}P_{PS} + N_{RF}P_C + N_{ANT}P_{SP} \\
\end{split}
\end{equation}
\normalsize
and therefore $b^*$ and $B^*$ for HBF and DBF can be calculated as 
\small 
\begin{equation}
\begin{split}
& R = \frac{N_{ANT}(N_{RF}P_{PS} + P_{SP}) + N_{RF}P_C - (N_{ANT} - N_{RF})P_{RF}}{2(N_{ANT} - N_{RF})cB} \\
& b^* = \lfloor log_2(R)\rfloor
\end{split}
\label{Eq:HDBF_b}
\end{equation}
\begin{equation}
 B^* = \frac{N_{ANT}(N_{RF}P_{PS} + P_{SP}) + N_{RF}P_C - (N_{ANT} - N_{RF})P_{RF}}{2(N_{ANT} - N_{RF})cR} 
\label{Eq:HDBF_B}
\end{equation}
\normalsize
where $\lfloor x \rfloor$ represents the floor of the variable $x$, i.e., the largest integer $\leq x$.  
Eqs. (\ref{Eq:HDBF_b}) and (\ref{Eq:HDBF_B}) hold for $N_{RF} < N_{ANT}$.
Now if $N_{ANT} \rightarrow \infty$, $b^*$ and $B^*$ are given by
\small
\begin{equation}
 b^* = \bigg{\lfloor} log_2(\frac{N_{RF}P_{PS} + P_{SP} - P_{RF}}{2cB})\bigg{\rfloor}
\label{Eq:HDBF_infb}
\end{equation}
\begin{equation}
 B^* = \frac{N_{RF}P_{PS} + P_{SP} - P_{RF}}{2cR}
 \label{Eq:HDBF_infB}
\end{equation}
\normalsize
\begin{figure}
        \centering
        \psfrag{NRF8}{\back\back\back \ $N_{RF} = N_{ANT} / 8 $}
        \psfrag{Tx-Rx Distance }{\back\ $Tx-Rx Distance$}
        \psfrag{Bandwidth}{\ \ \ \ $B$}
        \psfrag{Number of bits}{\ \ \ \ \ \ \ $b^*$}
        \scalebox{1}[0.70]{\includegraphics[width=\columnwidth]{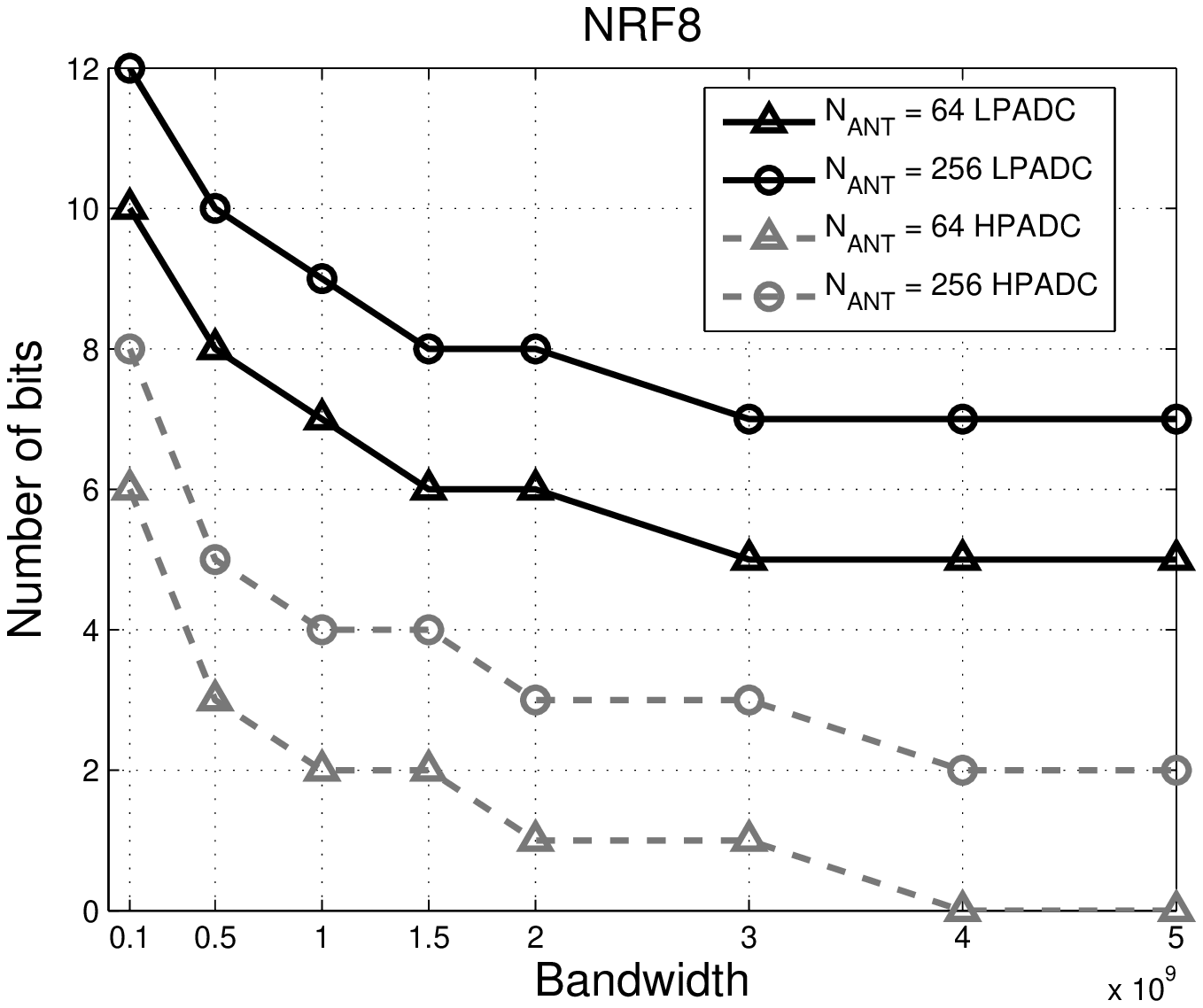}}\vspace{-4mm}
        \caption{\protect\renewcommand{\baselinestretch}{1.25}\footnotesize $b^*$ vs $B$ for HBF vs DBF considering both HPADC and LPADC.  }
\label{fig:HDBF_Opt_b2}
\end{figure}

\begin{figure}
        \centering
        \psfrag{NRF8}{\back\back $N_{RF} = N_{ANT} / 8$}
        \psfrag{Tx-Rx Distance }{\back\ $Tx-Rx Distance$}
        \psfrag{bandwidth}{\ \ \ \ \ $B^*$}
        \psfrag{bits}{\ $b$}
        \scalebox{1}[0.70]{\includegraphics[width=\columnwidth]{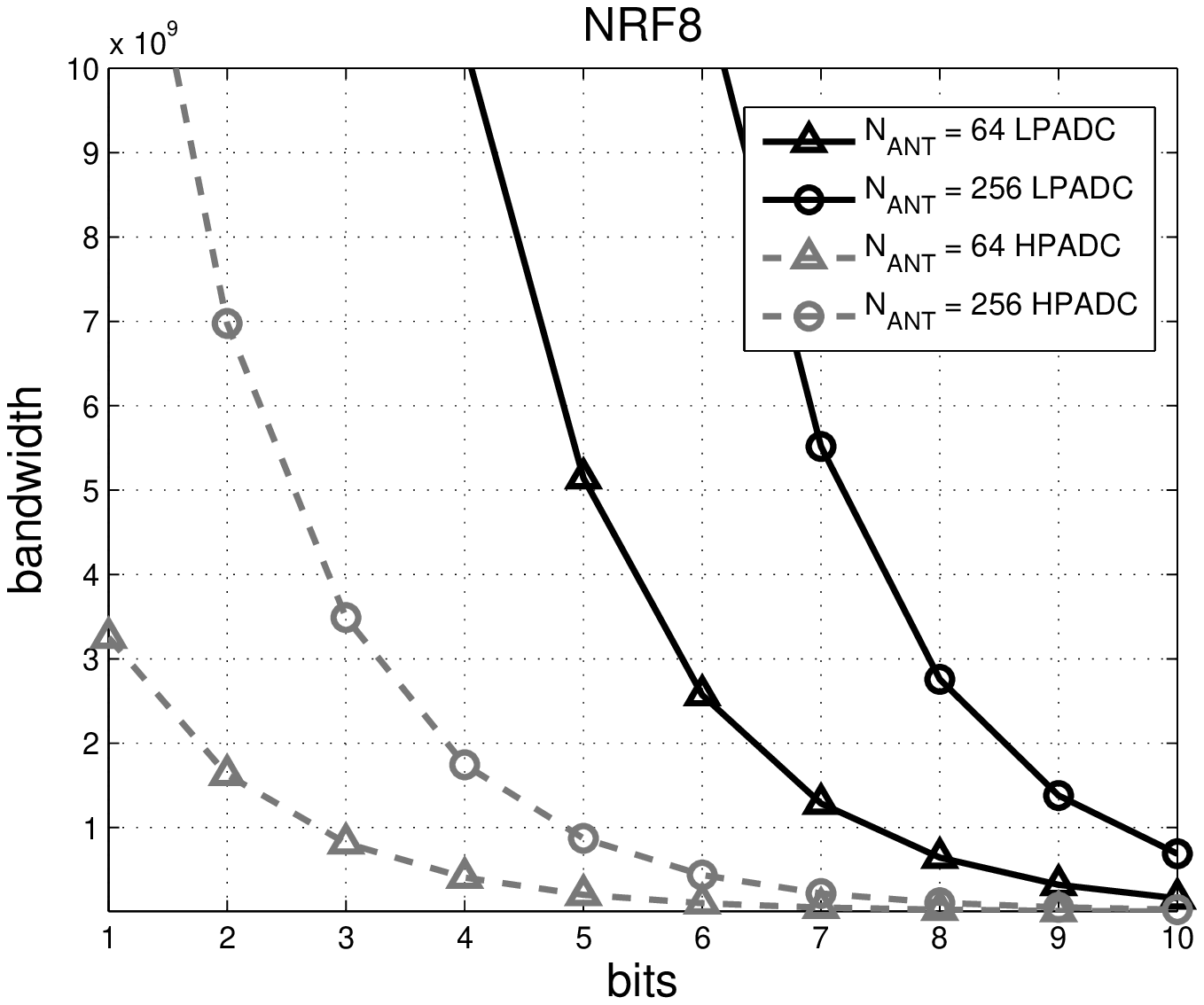}}\vspace{-4mm}
        \caption{\protect\renewcommand{\baselinestretch}{1.25}\footnotesize $B^*$ vs $b$ for HBF vs DBF considering both HPADC and LPADC.  }
\label{fig:HDBF_Opt_B2}
\end{figure}

Eqs. (\ref{Eq:HDBF_infb}) and (\ref{Eq:HDBF_infB}) show that, for a large number of antennas, the values of $b^*$ and $B^*$ for DBF are inversely related to $B$ and $b$, respectively, and directly related to $N_{RF}$.
Moreover, for constant $P_{PS}$, $P_{RF}$, $P_{SP}$, $c$, $N_{RF}$ and $b$ or $B$, Eqs. (\ref{Eq:HDBF_infb}) and (\ref{Eq:HDBF_infB}) also provide a lower bound for $b^*$ and $B^*$, respectively, for any $N_{ANT}$. 
In addition, note that if $N_{RF}$ increases in proportion to $N_{ANT}$, then the values of $b^*$ and $B^*$ will increase with an increase in $N_{ANT}$.

A detailed analysis is shown in Figures \ref{fig:HDBF_Opt_b2} and \ref{fig:HDBF_Opt_B2}, where we assume $N_{RF} = N_{ANT}/8$. 
This increase in $N_{RF}$ in proportion to $N_{ANT}$ ensures that the system performance comparison among HBF and DBF remains the same.
In Figures \ref{fig:HDBF_Opt_b2} and \ref{fig:HDBF_Opt_B2}, $b^*$ and $B^*$ are plotted, respectively, for different values of $N_{ANT}$ while considering both LPADC and HPADC.
As expected, $b^*$ and $B^*$ for HBF and DBF are higher for LPADC than for HPADC. 
Moreover, $b^*$ and $B^*$ increase with an increase in $N_{ANT}$.
This is due to the dependence of $N_{RF}$ on $N_{ANT}$.
This shows that large antenna systems with DBF based receivers can take advantage of a large bandwidth and/or a higher number of ADC bits while keeping the power consumption similar to that of HBF.
Moreover, by increasing the $N_{ANT}/N_{RF}$ ratio, the number of RF chains for a fixed $N_{ANT}$ decreases and therefore $P_{Tot}^{HBF}$ decreases.
This decrease in $P_{Tot}^{HBF}$ results in a reduction of $b^*$ and $B^*$.

Moreover, note that for $B = 1.5$ GHz and $N_{ANT} = 256$, the DBF receiver outperforms HBF when using ADCs with up to $b = 4$ and $b = 8$ bits for HPADC and LPADC, respectively.
These values of $b$ are large enough not to result in any significant SNR loss compared to a high resolution ADC, as discussed in Section \ref{sec:bits_SNR}.
With these configurations, DBF may be a preferable option than HBF for mmW receiver design.
It is also important to note that with an increase in $N_{ANT}$ it is very difficult to acquire the complete channel state information with a fully digital architecture (i.e., DBF) as it requires a very high complexity receiver design, whereas HBF decreases this complexity but at the cost of lower flexibility.
Therefore, the choice between DBF and HBF may also depend on flexibility and/or complexity, which are directly related to the application requirements. 
The detailed study of these complexity/flexibility issues is left as a future work.
Finally, as $B$ increases $b^*$ decreases (Figure \ref{fig:HDBF_Opt_b2}) and similarly $B^*$ decreases with an increase in $b$ (Figure \ref{fig:HDBF_Opt_B2}). 
This shows a trade off between the choice of $b$ and $B$, which means that for a high bandwidth receiver design $b$ should be reduced to keep the total power consumption within the required budget and vice versa.

\section{ADC Bits vs SNR}
\label{sec:bits_SNR}
We now extend our analysis to study how a change in the number of ADC bits affects the SNR.
We consider an AQNM based model as in \cite{OrhanER15_PowerCons}.
For this model, the effective SNR ($\gamma_{ef}$) is defined as \cite{Barati_IBF} 
\begin{equation}
 \gamma_{ef} = \frac{(1-\eta )\gamma}{1+\eta \gamma}
 \label{Eq:SNReff}
\end{equation}
where $\gamma$ and $\gamma_{ef}$ represent the SNR of a high resolution ADC and the effective SNR of a low resolution ADC, respectively, and $\eta$ is the inverse of the signal-to-quantization-noise ratio of the ADC, which 
depends on the quantizer design, the input distribution and the number of bits $b$.
For a gaussian input distribution, the values of $\eta$ for $b \leq 5$ are listed in Table \ref{tab:etavsb}, and for $b > 5$ can be approximated by  $\eta = \frac{\pi \sqrt{3}}{2} 2^{-2b}$ \cite{Fan_ULRate_LowADC15}.

\begin{figure}
        \centering
        \psfrag{Number of bits}{\ \ \ \ \ \ \ $b$}
        \psfrag{Effective SNR}{\ \ \ \ \ $\gamma_{ef} $ (dB)}
        \psfrag{SNR = -10 dB p p}{\scriptsize $\gamma = -10$ dB}
        \psfrag{SNR = 0 dB}{\scriptsize $\gamma = 0$ dB}
        \psfrag{SNR = 10 dB}{\scriptsize $\gamma = 10$ dB}
        \psfrag{SNR = 20 dB}{\scriptsize $\gamma = 20$ dB}
        \scalebox{0.9}[0.60]{\includegraphics[width=\columnwidth]{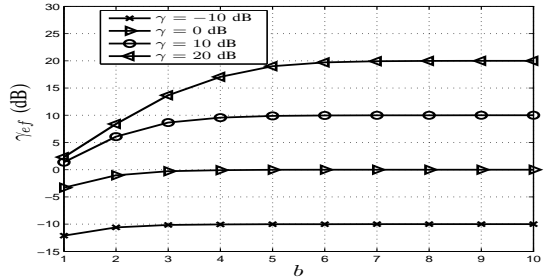}}\vspace{-4mm}
        \caption{\protect\renewcommand{\baselinestretch}{1.25}\footnotesize Change in effective SNR $\gamma_{ef}$ with a variation in the number of ADC bits $b$. }
\label{fig:SNReff}
\end{figure}

Figure \ref{fig:SNReff} shows how $\gamma_{ef}$ varies with the number of ADC bits. 
Each curve corresponds to a different value of $\gamma$, where $\gamma$ is varied from $-10$ to $20$ dB.
The results show that there is a number of bits $b_m$ after which any further increase in $b$ will not result in a significant increase in $\gamma_{ef}$, as $\gamma_{ef} \simeq \gamma$ for $b = b_m$.
Moreover, they also show that $b_m$ increases as we move from low to high SNR regime.
For instance, $b_m$ for $\eta = -10$ dB is 3 bits, whereas for $\eta = 20$ dB it goes up to 6 bits.
Therefore, the SNR regime which identifies $b_m$ is also directly related to the choice of the appropriate beamforming schemes.

  \begin{table}
     \centering
     \caption{$\eta$ for different values of $b$ \cite{Fan_ULRate_LowADC15}}
     \begin{tabular}{|c||c|c|c|c|c|}
         \hline
         $b$ &  1 & 2 & 3 & 4 & 5\\ 
         \hline
         $\eta$  & 0.3634 & 0.1175 & 0.03454 & 0.009497 & 0.002499 \\
         \hline
     \end{tabular}
     \label{tab:etavsb}
 \end{table}

\subsection{$P_{Tot}$ vs $\gamma_{ef}$ Comparison}
\label{ssec:PvsSef}
To summarize the analysis and to better identify the appropriate configuration for DBF, we now show a comparison between $P_{Tot}$ and $\gamma_{ef}$ for different values of $b$.
In particular, we combine the results provided in Figure \ref{fig:SNReff} with those of Figures \ref{fig:AHDBF_100M} and \ref{fig:AHDBF_1G}, and plot the effective SNR performance $\gamma_{ef}$ vs. the total power consumption, $P_{Tot}$ (where the different points on the curves correspond to different values of $b$), thereby highlighting the tradeoff between the power spent and the performance achieved.

\begin{figure}
        \centering
        \psfrag{NRFNANTby16}{\back $N_{RF} = N_{ANT} / 16 $}
        \psfrag{Ptot}{\back\back $P_{Tot}$ (Watts)}
        \psfrag{Effective SNR}{\ \ \ \ \ \ $\gamma_{ef}$ dB}
        \psfrag{SNR = 10 dB, Nrf = 4}{SNR = 10 dB, $N_{RF}$ = 4}
        \scalebox{0.9}[0.65]{\includegraphics[width=\columnwidth]{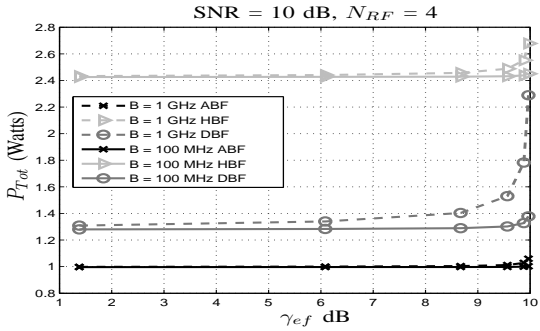}}\vspace{-4mm}
        \caption{\protect\renewcommand{\baselinestretch}{1.25}\footnotesize $P_{Tot}$ vs $\gamma_{ef}$ for LPADC model.}
\label{fig:PtotvsSNRef_NYU}
\end{figure}

\begin{figure}
        \centering
        \psfrag{NRFNANTby16}{\back $N_{RF} = N_{ANT} / 16 $}
        \psfrag{Ptot}{\back\back $P_{Tot}$ (Watts)}
        \psfrag{Effective SNR}{\ \ \ \ \ \ $\gamma_{ef}$ dB}
        \psfrag{SNR = 10 dB, Nrf = 4}{SNR = 10 dB, $N_{RF}$ = 4}
        \scalebox{0.9}[0.65]{\includegraphics[width=\columnwidth]{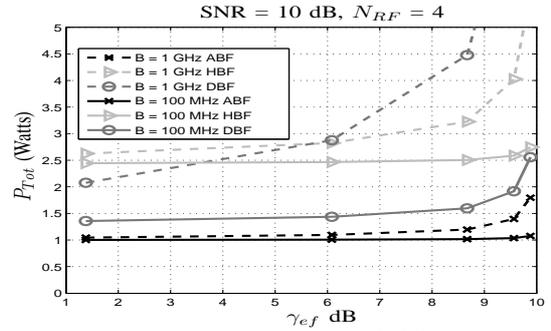}}\vspace{-4mm}
        \caption{\protect\renewcommand{\baselinestretch}{1.25}\footnotesize $P_{Tot}$ vs $\gamma_{ef}$ for HPADC model.}
\label{fig:PtotvsSNRef_Ahmed}
\end{figure}

Figures \ref{fig:PtotvsSNRef_NYU} and \ref{fig:PtotvsSNRef_Ahmed} show a comparison between $P_{Tot}$ and $\gamma_{ef}$ for a low and high power consumption ADC model, respectively.
Results are obtained for $\gamma = 10$ dB, $N_{RF} = 4$, $N_{ANT} = 16$ and for $B = 1$ GHz and $B = 100$ MHz. 
Markers on each curve correspond to increasing values of $b$ when going left to right, where $b$ varies from 1 to 6 and from 1 to 5 in Figures \ref{fig:PtotvsSNRef_NYU} and \ref{fig:PtotvsSNRef_Ahmed}, respectively.
Results show that an appropriate configuration for DBF is directly related to the ADC power consumption values.
For instance, with LPADC (Figure \ref{fig:PtotvsSNRef_NYU}), DBF has lower power consumption than HBF even up to 6 bits, for both $B = 1$ GHz and $B = 100$ MHz. 
However, with HPADC, DBF has a similar power consumption to HBF up to 5 and 2 bits for $B = 100$ MHz and $B = 1$ GHz, respectively, and rapidly becomes worse as the number of bits is increased. 
Note that, with 5 bits, $\gamma_{ef}$ is almost equal to a value which it can attain with infinite $b$. 
Therefore, DBF with LPADC is an attractive choice, and preferable to HBF for both $B = 100$ MHz and $B = 1$ GHz, whereas with HPADC, DBF is a feasible choice only for $B= 100$ MHz.

The results also show that ABF is always a better option from a power consumption perspective. 
However, note that DBF for both $B = 100$ MHz and $B = 1$ GHz with LPADC and for $B = 100$ MHz with HPADC results in approximately $30 \%$  more power consumption in comparison to ABF, and this percentage increases with an increase in $B$ or $N_{ANT}$.
Therefore, for receivers with a relatively small number of antennas and low bandwidth requirements, DBF may be a preferable choice as, for a limited increase in the total power consumption, it provides significant advantages in terms of flexibility, thanks to digital processing.
\section{Discussion}
\label{sec:Dis}
In this section, we discuss the choice of an appropriate beamforming scheme from the device (the MS or the BS) and the communication signal (control plane or data plane) perspective, as a function of the typical parameters of each configuration.
We discuss the choice of the beamforming scheme for the device and the communication signal separately.
In the former, we identify an appropriate beamforming scheme both at the MS and at the BS, focusing on their different form factor and application requirements.
In the latter, we identify the preferable beamforming scheme focusing on the different bandwidth requirements for control plane (CP) and data plane (DP) communication\footnote{Note that all these comparisons follow from the power consumption values of the receiver components mentioned in Section \ref{Sec:Sys_mod}, and therefore the desirable configuration of any beamforming scheme may vary with the change in the component's power consumption values. However, the general trend among the different beamforming schemes would remain the same, and the corresponding numerical values can be easily derived from our general analysis in Section \ref{Sec:Sys_mod}.}.
\subsection{Optimal Beamforming Scheme at the MS and at the BS}     
\label{ssec:opt_BF_BSMS}
The MS and the BS can accommodate different numbers of antennas and have different application requirements.
The MS can accommodate only a small number of antennas due to its small form factor and has simple application requirements, whereas the BS can accommodate a much higher number of antennas and has typically more advanced application requirements such as to ensure multi-user communication, etc.

\paragraph{Mobile Station} To ensure the constraints of a limited power budget and a small form factor, we assume that the MS can have at most 16 antennas (as assumed in most works).
It can be seen from Figures \ref{fig:PtotvsSNRef_NYU} 
 and \ref{fig:PtotvsSNRef_Ahmed} that with $N_{ANT} = 16$, the appropriate beamforming schemes are ABF and DBF. 
The choice between these schemes is application dependent.
For instance, during initial cell search, where the MS has to look in different angular directions to receive the synchronization signals, the formation of multiple simultaneous beams can be advantageous and therefore justify the additional power consumption.
In this case, DBF may in fact be a preferable choice than ABF as it allows to form multiple simultaneous beams which result in a lower search delay and in a reduced energy consumption, whereas  
with an ABF based receiver the MS has to look in all angular directions sequentially to identify the desired BS, which will increase the initial cell search delay and the total energy consumption. 
However, if the desired beamforming direction is already available and the advantages of DBF are not required (e.g., as in context information based schemes \cite{CaponeFS15_CI_BF}, \cite{myCI_PSN16}), then ABF can be a better option.


\paragraph{Base Station} The base station has to  simultaneously serve multiple MSs and, in contrast to the MS, can accommodate higher $N_{ANT}$ and has a much higher power available.
For analysis, we set the minimum $N_{ANT} = 64$ for the BS receiver design.
Moreover, to fulfill the requirement of serving multiple MSs, we primarily focus on the comparison of HBF and DBF, as with ABF at a particular instant a BS can communicate with only a single or a limited number of MSs. 

As discussed in Section \ref{Sec:Sys_mod}, the choice of the appropriate beamforming scheme between HBF and DBF depends not only on $N_{ANT}$, $b$ and $B$, but also on $N_{RF}$. 
As shown in Figure \ref{fig:HDBF_Opt_b2}, with LPADC model and low $N_{RF}$, DBF has a similar power consumption to HBF up to 6 bits even with $B = 2$ GHz, which makes DBF an appropriate choice with LPADC model, whereas in case of HPADC the resultant $b^*$ or $B^*$ for DBF is relatively low. 
Moreover, HBF generally allows to simultaneously communicate in only $N_{RF}$ different directions, whereas DBF with the same number of antennas can cover a much higher angular space and therefore can communicate with a larger number of MSs.
With this difference, 
and considering an interference free scenario, DBF can result in much higher capacity as compared to HBF.
However, to get the capacity with HBF similar to DBF and keeping $N_{ANT}$ fixed, we need to increase the number of RF chains which then results in an increase in $P_{Tot}^{HBF}$, which corresponds to an increase in $B^*$ and $b^*$, which may make DBF a more preferable design choice than HBF even with HPADC.

\subsection{Optimal Beamforming Scheme for Control and Data}     
\label{ssec:opt_BF_CPDP} 
Typically, the control plane (CP) and the data plane (DP) have different data rate requirements.
The CP has a low data rate requirement, which corresponds to a lower bandwidth, whereas the DP requires higher $B$ to support high data rate.
From the MS perspective, a power efficient receiver may require separate beamforming schemes for CP and DP signaling (Figures \ref{fig:PtotvsSNRef_NYU} and \ref{fig:PtotvsSNRef_Ahmed}).
For instance, to reduce the initial cell search delay during CP signaling and based on the low $B$ requirement for CP, DBF can be a preferable choice even with HPADC.
However, when the beamforming direction is already established and under the high data rate requirements of the DP, ABF may be a valid choice, especially at the MS side. 
On the other hand, for BSs that have to support more advanced applications and to support many users simultaneously, DBF may be a preferable choice for both CP and DP signaling.

\enlargethispage{-8mm}

\section{Conclusion}
\label{sec:Con}
In this work, we argued that a comparison of different beamforming schemes only based on ADC power consumption does not give full insight, and the results can be quite different when the total power consumption of the mmW receiver is considered. 
Based on power consumption analysis, we showed that there are cases in which DBF can still be a viable option compared to the other beamforming schemes. 
In particular, considering both a high power and a low power ADC for a mmW receiver model, we showed that for a certain range of $N_{ANT}$, $b$ and $B$, DBF may result in a lower power consumption than HBF. 
The results also showed that with a low number of antennas (e.g.,  as in case of the MS), DBF power consumption may be comparable to (or only slightly higher than) ABF power consumption.
Moreover, with SNR as a figure of merit, we showed that the number of ADC bits for DBF which results in a similar power consumption as in HBF is large enough to avoid any loss in SNR. 

In the future, we will study how the system capacity for different beamforming schemes varies with a change in the number of ADC bits and in the bandwidth.

\bibliographystyle{IEEEtran}
\bibliography{IEEEabrv,biblio,refen}

\end{document}